\documentclass[aps,prl,twocolumn,amsmath,amssymb,superscriptaddress]{revtex4}

\usepackage{amssymb}
\usepackage{amsmath}
\usepackage{graphicx}
\usepackage{subfigure}
\usepackage{textcomp}
\usepackage{color}
\usepackage{amsfonts}
\usepackage{bbold}
\usepackage{dsfont}
\usepackage{epsfig}
\usepackage{hyperref}

\begin{document}

\title{Assessment of density functional theory for iron(II) molecules across the spin-crossover transition}

\author{A. Droghetti}
\affiliation{School of Physics and CRANN, Trinity College, Dublin 2, Ireland}
\author{D. Alf\`e}
\affiliation{London Centre for Nanotechnology, Dept. of Earth Sciences, 
Dept. of Physics and Astronomy, University College London ,
Gower Street, London, WC1E 6BT,  UK}
\author{S. Sanvito}
\affiliation{School of Physics and CRANN, Trinity College, Dublin 2, Ireland}

\date{\today}

\begin{abstract}
Octahedral Fe$^{2+}$ molecules are particularly interesting as they often exhibit a spin-crossover transition. In spite 
of the many efforts aimed at assessing the performances of density functional theory for such systems, an 
exchange-correlation functional able to account accurately for the energetic of the various possible spin-states has 
not been identified yet. Here we critically discuss the issues related to the theoretical description of this class of molecules 
from first principles. In particular we present a comparison between different density functionals for four ions, namely 
[Fe(H$_2$O)$_6$]$^{2+}$, [Fe(NH$_3$)$_6$]$^{2+}$, [Fe(NCH)$_6$]$^{2+}$ and [Fe(CO)$_6$]$^{2+}$. 
These are characterized by different ligand-field splittings and ground state spin multiplicities. Since no experimental 
data are available for the gas phase, the density functional theory results are benchmarked against those obtained 
with diffusion Monte Carlo, one of the most accurate methods available to compute ground state total energies of 
quantum systems. On the one hand, we show that most of the functionals considered provide a good description of 
the geometry and of the shape of the potential energy surfaces. On the other hand, the same functionals fail badly in 
predicting the energy differences between the various spin states. In the case of [Fe(H$_2$O)$_6$]$^{2+}$, 
[Fe(NH$_3$)$_6$]$^{2+}$, [Fe(NCH)$_6$]$^{2+}$, this failure is related to the drastic underestimation of the exchange 
energy. Therefore quite accurate results can be achieved with hybrid functionals including about $50\%$ of Hartree-Fock 
exchange. In contrast, in the case of [Fe(CO)$_6$]$^{2+}$, the failure is likely to be caused by the multiconfigurational 
character of the ground state wave-function and no suitable exchange and correlation functional has been identified.
\end{abstract}
\maketitle

\section{Introduction}
\begin{figure}[h!]
\centering \includegraphics[scale=0.6,clip=true]{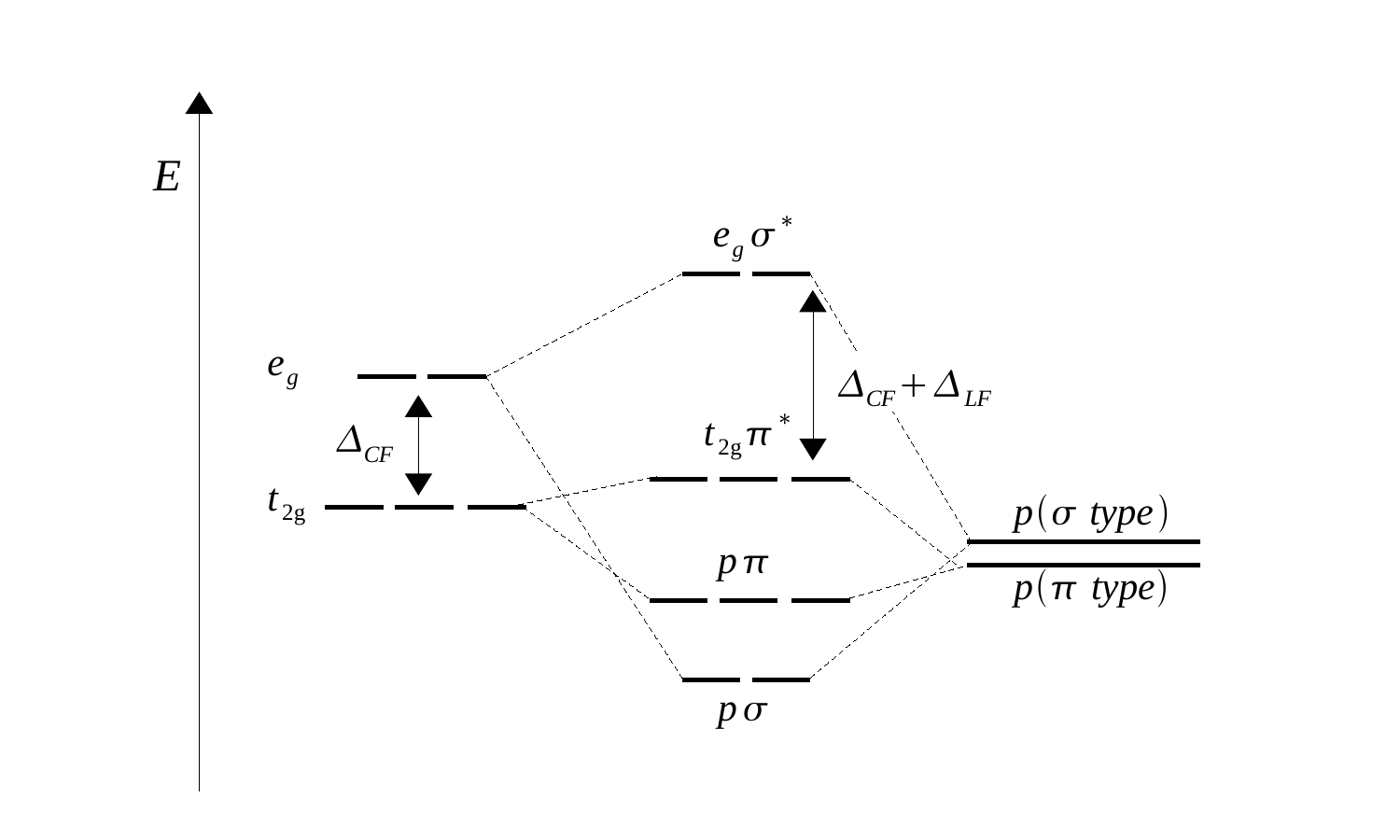}
\caption{Energy level diagram for an octahedrally coordinated transition metal (TM) ion. In crystal field theory, the $3d$ 
orbitals of the TM ion have an energy split $\Delta_\mathrm{CF}$ due to the electrostatic interaction with the ligands. 
In contrast, in ligand field theory the $3d$ orbitals of the TM ion form covalent bonds with the ligands. In this 
diagram we assume that each ligand contributes three $p$-orbitals, one with the positive lobe pointing toward the TM 
ion, $\sigma$-type, and two with the lobes perpendicular to it, $\pi$-type, (note that the $\sigma$- and $\pi$-type $p$ 
orbitals are degenerate but here they are plotted slightly separated for better display). The  $\pi$-type $p$ orbitals couple 
with the TM $t_{2g}$ states, while the $\sigma$-type $p$ orbitals couple with the $e_g$. Since the $\pi$ interaction is 
weaker than the $\sigma$ one, the antibonding $t_{2g} \pi^*$ orbitals lie lower in energy than the  $e_{g} \sigma^*$ ones. 
The energy splitting between the $t_{2g} \pi^*$ and the $e_{g} \sigma^*$ orbitals is indicated by 
$\Delta_\mathrm{LF}+\Delta_\mathrm{CF}$.}\label{ligand_field}
\end{figure}

Among the transition metal complexes, octahedral $3d^6$ Fe$^{2+}$ molecules are systems of particular interest. In fact 
they often undergo the spin-crossover (SC) transition \cite{SC_molecules}. In their most common form the molecules ground 
state is a spin singlet ($S=0$), with the Fe six $3d$ electrons paired up in the $t_{2g}\pi^*$ antibonding orbitals. Upon
increasing the temperature, the high spin quintet state ($S=2$), in which two electrons are promoted from the $t_{2g}\pi^*$ 
to the $e_g\sigma^*$ orbitals, becomes thermodynamically more stable (see Fig. \ref{ligand_field} for the molecular
orbital diagram). The SC transition is entropy driven and it is regulated by the difference in the Gibbs free energy of the 
HS and LS states,
\begin{equation}
\Delta G=G_\mathrm{HS}-G_\mathrm{LS}=\Delta H-T \Delta S\,.\label{deltaG}
\end{equation}
Here $\Delta H =H_\mathrm{HS}-H_\mathrm{LS}$ and $\Delta S =S_\mathrm{HS}-S_\mathrm{LS}$ indicate 
respectively the enthalpy and the entropy variation (note that for $\Delta G>0$ the LS configuration is more 
thermodynamically stable than the HS one). For SC molecules $\Delta H>0$, but also $S_\mathrm{HS}>S_\mathrm{LS}$,
i.e. $\Delta S>0$. Hence, for large enough temperatures ($T>T_\mathrm{c}=\Delta H/\Delta S$), the entropic term 
dominates over the enthalpic one and the molecules transit from the LS to the HS configuration. There are two contributions 
to the entropy: the first is provided by the spin and the second by the molecule vibrations. In fact, the transfer of two 
electrons to the  $e_g\sigma^*$ orbitals, which are more-antibonding than the $t_{2g}\pi^*$, weakens the chemical 
bond and produces a breathing of the metal ion coordination sphere. This results in a softening of the phonon modes and 
then an increase of the vibronic entropy.

The SC transition is usually reported either for molecules in solution or in single crystals and, depending on the strength 
and on the origin of the inter-molecular interactions, it can have cooperative nature and present a thermal hysteresis loop.
Interestingly, the transition can be also induced by illumination. This phenomenon is called LIESST effect 
(Light-Induced-Excited-Spin-State-Trapping) and it is explained through a mechanism involving several excited 
states \cite{Hauser}. The lifetime of the metastable HS state is usually very long at low temperature as the relaxation to 
the ground state is due to the weak electronic coupling between these states \cite{Buhks}. The opposite process, in 
which molecules populating the metastable HS state are brought back to their ground state, is also possible and it is 
called reverse LIESST effect.

Although, SC molecules have been traditionally studied for possible applications in storage and sensor devices \cite{SC_molecules,Real,Gamez}, they have recently emerged as promising materials for molecular 
spintronics \cite{Sanvito1,bogani,Sanvito2}. In fact, the electronic transport through these molecules has been 
predicted \cite{Aravena,Nadjib} and experimentally reported \cite{Takahashi,Alam,Prins,Ruben} to depend strongly on 
their spin state.  

Given such renewed interest in spin crossover compounds there is also a growing fundamental effort in modeling
their properties. In this respect one aims at using an electronic structure theory, which is at the same time accurate 
and scalable. Accuracy is needed for reliable predictions of the spin crossover temperature, while
scalability is required by the size of the typical molecules. This becomes particularly crucial for molecules in 
in crystals and when deposited on metallic surfaces, since the typical simulation cells are large. Density functional theory (DFT) is in principle
both scalable and accurate, but to date it is completely unclear how it does perform relatively to this problem. 

In this paper we investigate the performance of DFT against four model Fe$^{2+}$ ions, namely 
[Fe(H$_2$O)$_6$]$^{2+}$, [Fe(NH$_3$)$_6$]$^{2+}$, [Fe(NCH)$_6$]$^{2+}$ and [Fe(CO)$_6$]$^{2+}$. In particular
we answer to the question on which exchange-correlation functional performs best. As no experimental data
are available for these ions we benchmark against diffusion Quantum Monte Carlo (QMC) results. Note that often 
experimental data are difficult to compare with microscopic theory, so that our work provides a rigorous benchmark for 
the theory itself, which is probably more informative than a simple comparison to experiments. The paper is organized 
as follows. In the next section we present an overview of the problem and of the various known shortcomings of DFT, and we
discuss critically which elements one has to consider when comparing electronic structure data to experiments. 
Then we will provide some computational details and move to the results. First we will discuss our DFT calculations
for the four different ions and then we will compare them with the QMC ones. Finally we will conclude.

\begin{figure}[ht!]
\centering \includegraphics[scale=0.28,clip=true]{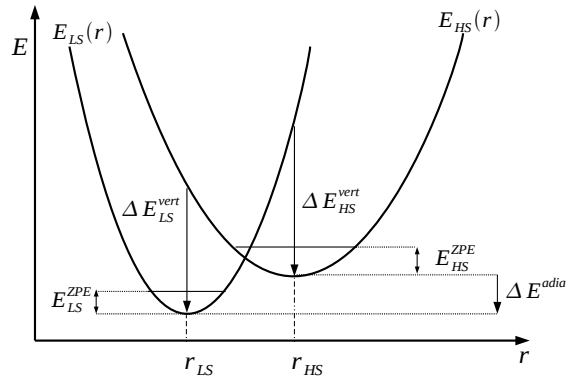}
\caption{Potential energy surface of the HS and LS state of a SC molecule. The collective coordinate $r$ represents all 
the 3$N$ nuclear coordinates of the molecule. The zero point phonon energies for the HS and LS state, 
$E_\mathrm{HS}^\mathrm{ZPE}$ and $E_\mathrm{LS}^\mathrm{ZPE}$, the adiabatic energy gap, 
$\Delta E^{\mathrm{adia}}$, and the vertical energy gaps, 
$\Delta E^{\mathrm{vert}}_\mathrm{LS}=\Delta E^{\mathrm{vert}}(r_\mathrm{LS})$ and 
$\Delta E^{\mathrm{vert}}_\mathrm{HS}=\Delta E^{\mathrm{vert}}(r_\mathrm{HS})$ are indicated.}\label{adia_pic}
\end{figure}

\section{Density functional theory for spin-crossover molecules: state of art}
When we consider a single molecule in vacuum at zero temperature $\Delta G$ coincides with the internal 
energy difference, which reads in the adiabatic approximation as
\begin{equation}
\Delta E=\Delta E^{\mathrm{adia}}+ \Delta E^\mathrm{ZPE}\,.\label{delta_E_int}
\end{equation}
Here, $\Delta E^\mathrm{ZPE}=E_\mathrm{HS}^\mathrm{ZPE}-E_\mathrm{LS}^\mathrm{ZPE}$ and 
$E_\mathrm{HS}^\mathrm{ZPE}$ ($E_\mathrm{LS}^\mathrm{ZPE}$) is the zero-point phonon energy of the 
HS (LS) state, while
\begin{equation}
\Delta E^{\mathrm{adia}}=E_\mathrm{HS}(r_\mathrm{HS})-E_\mathrm{LS}(r_\mathrm{LS})\label{delta_E_adia}
\end{equation}
is the adiabatic energy difference (often called ``adiabatic energy gap''). The collective coordinate $r$ represents 
the $3N$ nuclear coordinates of the molecule and the energies $E_\mathrm{HS}(r)$ and $E_\mathrm{LS}(r)$ define 
the potential energy surfaces (PESs) respectively of the HS and LS state (see Fig.~\ref{adia_pic}).
In addition to the adiabatic energy gap we can also define the vertical energy difference (``vertical energy 
gap'')\cite{foot1}
\begin{equation}
\Delta E^{\mathrm{vert}}(r)=E_\mathrm{HS}(r)-E_\mathrm{LS}(r)\label{vert_gap}
\end{equation}
and the difference of vertical energy gaps (DOG) \cite{Zein}:
\begin{equation}
\mathrm{DOG}(r_2,r_1)=\Delta E^{\mathrm{vert}}(r_2)-\Delta E^{\mathrm{vert}}(r_1)\,.\label{DOG}
\end{equation}
All of these quantities can be computed by using {\it ab-initio} electronic structure methods. As we mentioned in the introduction, DFT is 
always the preferred one. In fact SC molecules are composed of, at least, fifty atoms and a good balance 
between expected accuracy and computational cost is required. However, there are many issues connected 
to the DFT description of SC molecules, which either have not been properly addressed or have not found 
any satisfactory solution yet. Here we list some of them.
\begin{itemize}
\item[-] \textbf{The Functional dilemma}. For each Fe$^{2+}$ molecules, in general, every exchange-correlation 
functional returns a very different adiabatic energy gap (see for example references \cite{Swart_2,Zein}). These 
differences can be as large as few eV and different functionals do not sometimes even predict the same
$\Delta E^{\mathrm{adia}}$ sign. In a nutshell, no agreement around which functional performs best has been 
reached so far (the discussion below will explain how problematic is a direct assessment of the DFT results 
through a comparison with the experimental data). However, some general trends, which relate functionals belonging 
to the same ``class'' (or the same ``rung'' if we refer to the ``Jacob's ladder'' \cite{Perdew_Schmidt} classification 
scheme), have been identified. 

1) {\it the local density approximation (LDA) (first rung) tends to stabilize the LS state}. This is due to the 
underestimation of the exchange energy \cite{Engel}. In particular, the exchange keeps electrons of like-spin 
apart so that their Coulomb repulsion is reduced. Therefore, the exchange underestimation is accompanied 
by the overestimation of the Coulomb energy for two electrons of equal spin. This, in turns, leads to the stabilization 
of the LS state at the expense of the HS state. 

2) {\it the generalized gradient approximation (GGA) and the meta-GGAs (second and third rungs) \cite{Fiolhais} give 
results that depend on the specific compound and on the exact DFT conditions that each functional satisfies}. 
``Traditional'' GGAs, such as the Perdew-Burke-Ernzerhof (PBE) \cite{pbe} and the BLYP, which combines the 
Becke exchange \cite{Becke} with the Lee-Yang-Parr (LYP) correlation \cite{LYP}, reduce only slightly the LDA 
over-stabilization of the LS state. Therefore they do not represent a drastic improvement. In contrast, some more 
recent GGA functionals, such as the OLYP (combining the Handy and Cohen's OPTX exchange \cite{olyp} with 
the LYP correlation), have been claimed to perform rather well \cite{Reiher_3,Fouqueau_1, Fouqueau_2}. Among 
the meta-GGAs, the Van Voorhis-Scuseria exchange-correlation (VSXC) functional \cite{VSXC} was tested by 
Ganzenm\"uller et al. \cite{Reiher_3}, who concluded that it provided quite accurate results for single-point 
calculations, while it predicted artificially twisted structures.

3) {\it hybrid functionals (forth rung) tend to favor the HS with respect to LS one}. Reiher and 
co-workers \cite{Reiher_2,Reiher_1} argued that the amount of Hartree-Fock (HF) exchange in many 
hybrid functionals is too large to predict correct energy gaps. They then proposed a re-parametrization of 
the B3LYP functional \cite{b3lyp}, called B3LYP$^*$, which includes only $15\%$ of HF exchange (in contrast to 
the standard B3LYP $20\%$). Although B3LYP$^*$ is believed to give satisfactory results for several SC 
complexes, some studies suggested that a further reduction of the HF exchange could be needed in order to 
describe others Fe$^{2+}$ compounds \cite{Reiher_1, Reiher_3}. In contrast, the amount of HF exchange in 
B3LYP was judged insufficient for the small ions [Fe(H$_2$O)$_6$]$^{2+}$ and [Fe(NH$_3$)$_6$]$^{2+}$. 
For these systems it has been claimed that PBE0 \cite{Ernzerhof,Adamo}, which includes up to $25\%$ of HF 
exchange, gives more satisfactory results \cite{Fouqueau_1, Fouqueau_2}. In practice, for each compound, 
the amount of HF exchange can be varied to fit the desired values for the gaps but no ``universally good choice'' 
has been identified so far. The dependence of the adiabatic energy gap on the amount of HF and either 
of LDA (Slater) or GGA (B88 and OPTX) exchange is explained very clearly in the work of Ganzenm\"uller et al.
\cite{Reiher_3}. Finally it is important to remark that, even when a certain hybrid functional is found to return 
satisfactory energy gaps, it might not be the optimal functional to describe other properties of the molecule. 

\item[-] \textbf{Basis set}. DFT calculations for SC molecules are usually performed by using Quantum Chemistry 
codes, where the wave-function is expanded over either Gaussian (GTOs) or Slater-type orbitals (STOs)~\cite{Szabo}. 
In many cases, the values of the energy gaps depend substantially on the choice of the basis sets and on the spatial
extension of the local orbitals \cite{Swart}. Although this is a less severe problem compared to that of identifying the 
exchange-correlation functional, it must be kept into consideration. In principle, the use of plane-waves basis sets 
\cite{Martin}, instead of GTOs and STOs, could be a solution, but in practice plane-wave calculations are are 
computationally expensive because of the need of satisfying periodic boundary conditions. \\
A large number of plane-waves is usually needed as the electronic density is concentrated in a small fraction of 
the supercell volume. Furthermore very large supercells are typically required. This is due to the fact that SC complexes 
are often 2+ ions. Therefore a negatively charged background must be introduced in order to preserve the overall charge 
neutrality so that the total energy remains bound. The energy calculated in this way approaches then the one for 
an isolated system only in the limit of large supercell and, unfortunately, such convergence is slow. 
Although corrections to the expression of the computed energy have been proposed \cite{Leslie,Markov}, this effect can be 
properly accounted for only by considering large supercells and by performing multiple calculations for supercells of 
different sizes. 

\item[-] \textbf{Geometry optimization}. Each class (rung) of exchange-correlation functionals returns different metal-ligands 
bond-lengths. Usually, LDA gives shorter bonds than hybrids functionals, while standard GGA results are in between these 
two extremes \cite{foot2,me2}. Although these differences are usually quite small, less then $0.1$ \AA\ against an average 
bond-length of about $2$ \AA, they might strongly affect the electronic properties. Indeed a very delicate balance between 
ligand-field splitting and Hund's coupling establishes the spin state (see Fig.~\ref{ligand_field}). Then, small errors in the 
predicted geometries can drastically alter this balance. \\
Unfortunately, the quality of the DFT relaxation can not be easily assessed through a comparison with available experiments. 
In fact, while DFT calculations are usually carried out for molecules in the gas phase, the experimental geometries are 
obtained through X-ray measurements for crystals \cite{foot3}. As far as we know, no detailed DFT studies about SC 
molecules in the condensed phase have been published so far. Furthermore, such a study must face the additional 
difficulty of the need of accounting for inter-molecular interactions. These have usually dispersive nature and, therefore, 
they are either not described or badly-described by most of the popular functionals \cite{van_der_Waals}. 

\item[-] \textbf{Electrostatic contributions to the total energy}. 
SC molecules, which are usually in the 2+ charging state, are surrounded by counter-ions (for example [PF$_6$]$^{2-}$ 
or [BF$_4$]$^{2-}$). Because of the electrostatic potential generated by such counter-ions and by the other molecules in
the crystal, the total energy of a SC complex in the condensed phase differs from that in the gas phase. Furthermore, 
since at the SC phase transition there is a charge redistribution over the individual molecules and a lattice expansion, 
such electrostatic potential does not induce a simple rigid shift of the minima of the HS and LS PESs (compared to those 
calculated for the gas phase). The energy gap then turns out to be different for the same molecule in different phases 
\cite{Robert_1,Robert_2}. Unfortunately this effect is always neglected by DFT calculations, which aim at assessing the 
performances of DFT by using experimental data.

\item[-]\textbf{Finite-temperature effects}.
Special care should be taken in order to include finite-temperature effects when comparing DFT to experiments. In fact, 
at finite temperature instead of the adiabatic energy gap, the Gibbs free energy, Eq.~(\ref{deltaG}), must be computed. 
So far, this has been attempted only by Ganzenm\"uller et al. \cite{Reiher_3}. However, unfortunately, their calculations 
did not fully account for either the electrostatic contribution to the total energy or the effect of the periodic lattice on 
the molecular structure.   
\end{itemize}

This list clearly emphasizes how the main handicap in the theoretical description of SC complexes is related to 
the difficulties in assessing the performances of any given exchange-correlation functional. Any benchmark involving 
a comparison against experimental results is fated to fail, unless vibrational, environmental, crystallographic and 
finite-temperature effects are properly accounted for. However, this task is at present too demanding to be practically 
achievable. In contrast, as pointed out by Fouqueau et al. \cite{Fouqueau_1,Fouqueau_2}, the current best strategy 
consists in providing benchmark values for various interesting quantities through highly accurate {\it ab-initio} methods. 
These can then be compared with the DFT results in order to identify which functional performs better.

In some interesting works \cite{Fouqueau_1,Fouqueau_2,Pierloot1,Pierloot2,Ruben}, wave-function based methods 
were considered (see discussion below). However, unfortunately, the authors themselves admitted that their results 
were plagued by a number of systematic errors. These were ascribed to the too small basis set used for the 
Fe$^{2+}$ ion and to the fact that such computational methods describe exactly only for static electronic correlation, 
but do not include the dynamic one, which, however, can be partially included by perturbation theory.\\
Here we have chosen to employ as benchmark electronic structure theory 
diffusion Monte Carlo (DMC) \cite{Foulkes, casino,Gilliam}. This represents one of the most accurate computational 
techniques to calculate the energy of a quantum system and it is able to return up to $98\%$ of the correlation energy 
(including both static and dynamic contributions).

\begin{figure*}[ht!]
\centering
\includegraphics[scale=0.55, clip=true]{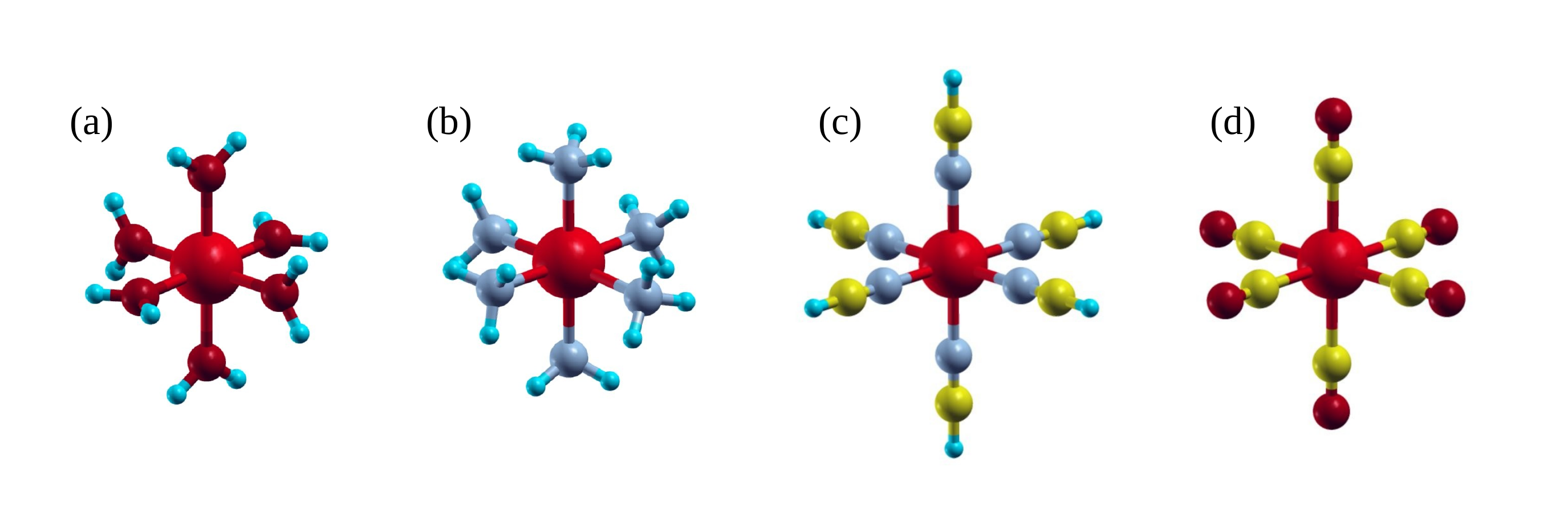}
\caption{(Color on line) The cations investigated in this work [Fe(H$_2$O)$_6$]$^{2+}$ (a), [Fe(NH$_3$)$_6$]$^{2+}$ (b), 
[Fe(NCH)$_6$]$^{2+}$ (c) and [Fe(CO)$_6$]$^{2+}$ (d). Color code: C=yellow, O=red (small sphere), Fe=red (large sphere), 
N=grey, H=blue.}\label{Fig_small_molecules}
\end{figure*}

In this work, we compare systematically several DMC results to those obtained with DFT for a few selected Fe$^{2+}$ 
complexes. Unfortunately, such a systematic investigation requires a large use of computational resources and, therefore, 
it can not be carried out for molecules composed of tens of atoms (such as the most typical SC complexes). We have then 
focused our attention on the following ions: [Fe(H$_2$O)$_6$]$^{2+}$, [Fe(NH$_3$)$_6$]$^{2+}$, [Fe(NCH)$_6$]$^{2+}$ 
and [Fe(CO)$_6$]$^{2+}$, which are shown in Fig. \ref{Fig_small_molecules}. These are small enough to allow several 
DMC calculations to be performed at a reasonable computational cost. Furthermore, and more importantly, the study of 
their electronic structure presents all the problems mentioned above so that our main conclusions can be extended to 
large SC molecules as well. Finally, according to the spectrochemical series \cite{Roberta_book}, these ions have a 
different ligand field splitting. The one of [Fe(H$_2$O)$_6$]$^{2+}$ is the smallest and that of [Fe(CO)$_6$]$^{2+}$ 
the largest. We then expect that Fe(H$_2$O)$_6$]$^{2+}$, [Fe(NH$_3$)$_6$]$^{2+}$ and [Fe(NCH)$_6$]$^{2+}$ have a 
HS ground state, while [Fe(CO)$_6$]$^{2+}$ a LS one. Therefore, our study scans through systems of different ground 
state spin multiplicity, it reveals general trends and it points to the systematic errors of each DFT functional.

\section{Computational details}
DFT calculations are performed with the NWCHEM package \cite{nwchem}. We use several functionals belonging to 
different ``classes'': 1) the default LDA of Vosko-Wilk-Nussair \cite{Vosko}, 2) the GGA BP86 functional, which combines 
the Becke88 exchange functional \cite{Becke} with the Predew86 correlation one \cite{Perdew86} and 3) the hybrid 
functionals B3LYP \cite{b3lyp}, PBE0 \cite{Ernzerhof,Adamo} and the Becke ``half and half'' (Becke-HH)~\cite{becke_HH}. 
These include respectivelly $20\%$, $25\%$ and $50\%$ of HF exchange (note that, in the NWCHEM package, only an 
approximate version of the HH-Becke functional is currently implemented \cite{NWCHEM_doc}).

We have chosen three different basis sets: 1) 6-31G* (basis set called A), 2) the LANL2DZ basis set and 
pseudopotential \cite{Hay} for Fe combined with the basis set 6-31++G** for all other atoms (basis set B) and 
3) the triple-zeta polarized basis set of Ahlrichs \cite{Ahlrichs} (basis set C). Geometry optimizations are performed both 
with and without specifying the molecule point group. The geometries optimized with these two strategies usually return 
consistent results with bond-lengths differences, which are within $\pm 0.005$\AA. We always check that the phonon 
frequencies are all real so that the final geometries correspond to stable energy minima. 

DMC calculations are performed by using the CASINO code \cite{casino}. The imaginary time evolution of the 
Schr\"odinger equation has been performed with the usual short time approximation and time-steps of various sizes 
are considered (typically $\Delta\tau=0.0125,0.005,0.001$ a.u.). Expect for few cases, the energy differences are usually found to be converged 
with respect to the time-step errors already for $\Delta\tau=0.0125$ a.u. Calculations are performed by using Dirac-Fock 
pseudopotentials \cite{Trail1,Trail2} with the  ``potential localization approximation'' (PLA) \cite{Mitas}. For [Fe(NCH)$_6$]$^{2+}$ 
DMC simulations with this approximation are found to be unstable as the number of walkers ``explodes''. Therefore we have 
used instead the ``T-move'' scheme \cite{Casula_1,Casula_2}, which eliminates the need of the PLA and treats the 
non-local part of the pseudopotential in a way consistent with the variational principle. The simulations then become more 
stable. The single-particle orbitals of the trial wave function are obtained through (LDA) DFT calculations performed with the 
plane-wave (PW) code QUANTUM ESPRESSO \cite{espresso}. The same pseudopotentials used for the DMC calculations 
are employed. The PW cutoff is fixed at $300$ Ry and the PW are re-expanded in terms of B-splines \cite{Dario}. The B-spline 
grid spacing is $a = \pi/ G_{\mathrm{max}}$, where $G_{\mathrm{max}}$ is the length of the largest vector employed in the 
PW calculations. Periodic boundary conditions are employed for the PW-DFT calculations and supercells as large as 
$40$~\AA~ are considered. In contrast, no periodic boundary conditions are imposed for the DMC simulations. 

DMC calculations are performed for the molecular geometries previously optimized by DFT. Therefore we can compare 
the DMC energies of molecular structures obtained by employing different functionals and basis sets. However, for each 
system, these energy differences are often smaller than the computed error bars. Only LDA systematically returns  
molecular geometries with much higher DMC energies than those obtained by using either GGA or hybrid functionals.  

\section{DFT Results} 
\subsection*{$[$Fe(H$_2$O)$_6$]$^{2+}$}
The lower energy geometry of [Fe(H$_2$O)$_6$]$^{2+}$ is found to have $C_i$ symmetry for both BP86 and 
hybrid functionals. This is consistent with the results of Pierloot et al. \cite{Pierloot}. In contrast, with LDA, we were able to obtain relaxed atomic positions for 
both the HS and LS states only by using the the basis set A and without specifying the molecule point group. 

As expected from our introductory discussion, the molecule in the LS state has metal-ligand bond-lengths shorter than 
those of the molecule in HS state (by about $7\%$). However, the details of the geometry depend on both the functional 
and the basis set. This can be clearly seen by inspecting Tab.~\ref{Tab_bond_1}, which reports a full list of the calculated 
Fe-O bond-lengths for both the HS and LS states. On the one hand, LDA overbinds the molecule as compared to GGA 
and hybrids. On the other hand, the basis set A tends to shrink the Fe-O bond-lengths when compared to the basis set 
B and C. Although the choice of basis set does not affect the bond-lengths as drastically as the functional does, it still  
influences the geometry greatly. The calculations performed with the basis set A return a quite large inclination (about 
$5$ degrees) of the O-Fe-O axis with respect to the $90$ degrees angle it forms with the equatorial plane of the molecule.
Furthermore, for basis set A and B, either the axial waters, which form the ligands, move ``in'' and the equatorial waters 
move ``out'' of their plane or viceversa. These results do not depend qualitatively on the functional. In contrast, the inclination 
of the O-Fe-O axis and the distortion of axial waters disappears when the calculations are carried out by using the basis set C.

Tab. \ref{Tab_dft_mine_1} shows our calculated values for the adiabatic energy gaps, where a positive (negative) energy 
means that the LS state has lower (higher) energy than that of the HS one. For each functional, our results are always in 
very close agreement with those obtained by other authors \cite{Fouqueau_2,Pierloot} (the results are presented in cm$^{-1}$ 
as well as in eV in order to allow for a better comparison with the various values found in literature). Here we can distinguish a 
clear trend, summarized by the series:
\begin{eqnarray}
&- \Delta E^{\mathrm{adia}}(\mathrm{LDA}) < - \Delta E^{\mathrm{adia}}(\mathrm{GGA})< -\Delta E^{\mathrm{adia}}(\mathrm{B3LYP}) \nonumber\\&<- \Delta E^{\mathrm{adia}}(\mathrm{PBE0})< -\Delta E^{\mathrm{adia}}(\mathrm{HH})\,.\label{func_series}
\end{eqnarray} 
This suggests that the calculated $\Delta E^{\mathrm{adia}}$ is strictly related to the amount of HF exchange incorporated into 
the given functional. By increasing such contribution, we systematically stabilize the HS configuration with respect to the LS one.
\begin{table}[h!]\centering
\begin{tabular}{lccc}\hline\hline
Functional   & Basis set & $d_\mathrm{LS}$ (\AA)& $d_\mathrm{HS}$ (\AA) \\ \hline\hline
LDA  & A& $1.917$ &$2.052,2.083,2.057$\\
BP86 &A & $1.985$ & $2.152,2.151,2.111$ \\
BP86 & B & $2.02$ &$2.174,2.164,2.132$ \\
BP86 & C & $2.01$& $2.161,2.155,2.125$  \\
B3LYP & A & $2.005$ & $2.152,2.152,2.111$\\
B3LYP & B & $2.003$ & $2.146,2.157,2.112$\\
B3LYP & C & $2.029$&  $2.172,2.16,2.137$ \\
PBE0  & A  & $1.99$ & $2.1,2.145,2.134$\\
PBE0 & B   &$2.013$ & $2.168,2.156,2.129$\\
PBE0 &  C & $2.008$ & $2.152,2.147,2.124$\\
HH   &  B & $2.010$ & $2.168,2.133,2.132$\\
HH  &  C  &  $2.008$ & $2.149, 2.131,2.128$\\
\hline\hline
\end{tabular}
\caption{Bond-lengths of [Fe(H$_2$O)$_6$]$^{2+}$ in the HS and LS state, as calculated with various functionals 
and basis sets. Note that the LDA calculations for the HS state did not coverge in the case of the basis set B and 
C.}\label{Tab_bond_1}
\end{table}
\begin{table}[h!]\centering
\begin{tabular}{lccc}\hline\hline
Functional   & Basis set & $\Delta E^{\mathrm{adia}}$ (cm$^{-1}$)& $\Delta E^{\mathrm{adia}}$ (eV) \\ \hline\hline
LDA  &A & $-3986$& $-0.4942$\\
BP86 &A & $-8989$ & $-1.1145$ \\
BP86 & B & $-8381$ & $-1.0391$\\
BP86 & C & $ -8400$& $-1.0415$\\
B3LYP & A & $-11589$ & $-1.4369$\\
B3LYP & B & $-11027$ & $-1.3672$\\
B3LYP & C &$-11045$ & $-1.3694$\\
PBE0  & A  & $-14670$ &$-1.8189$ \\
PBE0 & B   & $-15512$ & $-1.9233$\\
PBE0 &  C &  $-14045$&$-1.7414$\\
HH  & C & $-19620$ & $-2.4326$\\
HH  & C &  $-18223$& $-2.2594$\\ \hline\hline
\end{tabular}
\caption{Adiabatic energy gap, $\Delta E^{\mathrm{adia}}$, for the cation [Fe(H$_2$O)$_6$]$^{2+}$. The functional and 
the basis set used for the each calculation are indicated.}\label{Tab_dft_mine_1}
\end{table}
\begin{table}[h!]\centering
\begin{tabular}{lccc}\hline\hline
Functional   & Basis set & $d_\mathrm{LS}$ (\AA)& $d_\mathrm{HS}$ (\AA) \\ \hline\hline
LDA  &A &$1.942$ & $2.188,2.162,2.160$\\
LDA & C&  $1.995$&$ 2.204,2.201,2.214$\\
BP86 &A & $2.026$&$2.267,2.254,2.253$  \\
BP86 & B & $2.078$ & $2.30,2.295,2.274$\\
BP86 & C & $2.085$ & $2.279, 2.302,2.289$\\
B3LYP & A & $2.076$ & $2.281,2.281,2.275$\\
B3LYP & B & $2.114$ & $2.315,2.296,2.294$\\
B3LYP & C & $2.122$& $2.32,2.308,2.283$\\
PBE0  & A  & $2.05$  & $2.254,2.256,2.256$\\
PBE0 & B   & $2.082$ & $2.292,2.277,2.272$\\
PBE0 &  C & $2.093$ &$2.284,2.294,2.263$\\
HH  &  C  &  $2.11$  & $2.296,2.286,2.266$\\ \hline\hline
\end{tabular}
\caption{Bond-lengths of [Fe(NH$_3$)$_6$]$^{2+}$ in the HS and LS state, as calculated with various functionals and 
basis sets. Note that the LDA calculations for the HS state did not coverge in the case of basis set B.}  \label{Tab_bond_2}
\end{table}
\begin{table}[h!]\centering
\begin{tabular}{lccc}\hline\hline
Functional   & Basis set & $\Delta E^{\mathrm{adia}}$ (cm$^{-1}$)& $\Delta E^{\mathrm{adia}}$ (eV) \\ \hline\hline
LDA &  A&  $8937$  & $1.1081$\\
LDA &C & $7746$ & $0.9605$ \\
BP86 &A & $195$ & $0.0242$ \\
BP86 & B & $708$& $0.0878$\\ 
BP86 & C & $672$& $0.0834$\\
B3LYP & A & $-5312$ & $-0.6586$\\
B3LYP & B & $ -4007$& $-0.4969$ \\
B3LYP & C & $ -4738$& $-0.5874$\\
PBE0  & A  & $-7695$& $-0.9541$ \\
PBE0 & B   & $-7665$  & $-0.9504$\\
PBE0 &  C & $-7117$ & $-0.8825$\\
HH  &  C &  $-13556$ & $-1.68077$\\ \hline\hline
\end{tabular}
\caption{Adiabatic energy gap, $\Delta E^{\mathrm{adia}}$, for the cation [Fe(NH$_3$)$_6$]$^{2+}$. The functional and 
the basis set used for each calculation are indicated.}\label{Tab_dft_mine_2}
\end{table}

\subsection*{[Fe(NH$_3$)$_6$]$^{2+}$}
The optimized structure of the [Fe(NH$_3$)$_6$]$^{2+}$ ion, calculated either with BP86 or with hybrid functionals, has 
a $D_3$ symmetry for the LS state. This is further lowered to $C_2$ for the HS one. Our results are again consistent with 
those of  Pierloot et al. \cite{Pierloot}. Like in the case of [Fe(H$_2$O)$_6$]$^{2+}$, we were not able to find the relaxed 
atomic geometry with LDA. Even when the geometry optimization procedure converges, like in the case of the basis set A 
and C, the minimum is found to be unstable. This is indicated by the negative eigenvalues of some phonon modes. Nevertheless 
we report these results for completeness.

The molecule in LS state has shorter average Fe-ligand bond-lengths than the molecules in HS state (see Tab. \ref{Tab_bond_2}). 
In contrast to the case of [Fe(H$_2$O)$_6$]$^{2+}$, [Fe(NH$_3$)$_6$]$^{2+}$ does not show any strong deviation of the 
N-Fe-N axis with respect to the axis normal to the equatorial plane for any combination of functionals and basis sets.

Tab. \ref{Tab_dft_mine_2} shows several values for the adiabatic energy gaps. Once again these can be ordered according 
to the series (\ref{func_series}). Here, the LDA adiabatic energy gap indicates that [Fe(NH$_3$)$_6$]$^{2+}$ is LS. This result 
is even qualitatively incorrect as this cation is known to be stable in the HS state. BP86 also predicts the LS state to be the lowest 
in energy, although the value of $\Delta E^{\mathrm{adia}}$ is very small and probably very sensible to the exact details of 
the calculation. In fact, in contrast to our results, which agree with those in Ref.~\cite{Fouqueau_2}, Pierloot et al. \cite{Pierloot} 
obtained a negative value equal to about $-0.2$~eV. Finally hybrid functionals predict the ground state to be HS with the value 
of the gap being proportional to the amount of HF exchange included in the functional.
\begin{table}[ht!]\centering
\begin{tabular}{lccc}\hline\hline
Functional   & Basis set & $d_\mathrm{LS}$ (\AA)& $d_\mathrm{HS}$ (\AA) \\ \hline\hline
LDA  & C& $1.854$ &  $2.066,2.067,2.11$ \\
BP86 & C & $1.917$ & $2.171,2.171,2.155$ \\
B3LYP & C & $1.974$ & $2.206,2.201,2.201$\\
PBE0 &  C & $1.950$ & $2.194,2.181,2.181$\\
HH &  C &   $1.990$ & $2.20,2.196,2.196$ \\ \hline\hline
\end{tabular}
\caption{Bond-lengths of [Fe(NCH)$_6$]$^{2+}$ in the HS and LS state as calculated with various functionals and 
for the basis sets C.}\label{Tab_bond_3}
\end{table}
\begin{table}[ht!]\centering
\begin{tabular}{lccc}\hline\hline
Functional   & Basis set & $\Delta E^{\mathrm{adia}}$ (cm$^{-1}$)& $\Delta E^{\mathrm{adia}}$ (eV) \\ \hline\hline
LDA  &C & $19126.3$ & $2.37135$\\
BP86 & C & $8410.89$ &$1.04282$\\
B3LYP & C & $-1667.48$ & $-0.20674$ \\
PBE0 &  C & $-3544.58$&  $-0.43947$\\
HH  & C &    $-12029.62$  & $-1.49148$\\ \hline\hline
\end{tabular}
\caption{Adiabatic energy gap, $\Delta E^{\mathrm{adia}}$, for the cation [Fe(NCH)$_6$]$^{2+}$ calculated with various 
functionals and the basis sets C.}\label{Tab_dft_mine_3}
\end{table}

\subsection{[Fe(NCH)$_6$]$^{2+}$}
The results for the [Fe(H$_2$O)$_6$]$^{2+}$ and [Fe(NH$_3$)$_6$]$^{2+}$ ions demonstrate that, although a good choice 
of basis set might be important to predict accurate molecular structures, the estimated values of the adiabatic energy gaps 
depend mainly on the functional used. Indeed, for a given functional, two adiabatic gaps obtained with two different basis sets 
differ at most by a few tens of meV. This has to be compared with the differences in the values predicted by different functionals, 
which can be of several hundreds meV. For the ion [Fe(NCH)$_6$]$^{2+}$ we have then decided to compare only 
calculations performed using the basis set C, which typically gives us the lowest energy. 

$[$Fe(NCH)$_6$]$^{2+}$ has perfect octahedral symmetry in the LS state. In contrast, the structure of the HS state is predicted 
to have $D_{4h}$ symmetry by B3LYP and PBE0 and $C_i$ symmetry by BP86 and Becke-HH. Tab. \ref{Tab_dft_mine_3} 
displays the values of adiabatic energy gap calculated with each functional. Once again these can be ordered according to the 
series (\ref{func_series}). We find that the total energy of the LS state is at least $1$~eV lower than that of the HS state for both 
the LDA and BP86. In contrast PBE0 and B3LYP return the HS state as the most stable, but the absolute value of 
$\Delta E^{\mathrm{adia}}$ is only a few hundreds meV (note that our B3LYP adiabatic energy gap is consistent with that 
calculated by Bolvin \cite{Bolvin}). Finally, the Becke-HH predicts $\Delta E^{\mathrm{adia}}\approx -1.5$ eV. 
\begin{table}[ht!]\centering
\begin{tabular}{lccc}\hline\hline
Functional   & Basis set & $d_\mathrm{LS}$ (\AA)& $d_\mathrm{HS}$ (\AA) \\ \hline\hline
LDA  & C &   $1.848$ & $2.199,2.172,2.123$\\
BP86 & C & $1.900$ &  $2.226,2.331$\\
B3LYP & C & $1.948$ & $2.307,2.367$\\
PBE0 &  C &  $1.915$&  $2.276,2.345$\\
HH &  C &  $1.915$ & $2.322,2.329,2.366$\\ \hline\hline
\end{tabular}
\caption{Bond-lengths of [Fe(CO)$_6$]$^{2+}$ in the HS and LS state, as calculated with various functionals 
and for the basis set C.}\label{Tab_bond_4}
\end{table}
\begin{table}[ht!]\centering
\begin{tabular}{lccc}\hline\hline
Functional   & Basis set & $\Delta E^{\mathrm{adia}}$ (cm$^{-1}$)& $\Delta E^{\mathrm{adia}}$ (eV) \\ \hline\hline
LDA  & C &  $41148$  & $5.1017$\\
BP86 & C & $27575$ &$3.4189$\\
B3LYP & C & $10656$ & $1.32126$\\
PBE0 &  C & $10888$ & $1.3501$\\
HH &  C  & $5232$ &  $-0.6488$\\ \hline\hline
\end{tabular}
\caption{Adiabatic energy gap, $\Delta E^{\mathrm{adia}}$, for the cation [Fe(CO)$_6$]$^{2+}$ calculated with various 
functionals and the basis set C.}\label{Tab_dft_mine_4}
\end{table}

\subsection{[Fe(CO)$_6$]$^{2+}$}
The [Fe(CO)$_6$]$^{2+}$ ion has perfect octahedral symmetry in the LS state. This is then reduced to $D_{4h}$ in the 
HS (the metal-ligand bond-lengths are listed in Tab.~\ref{Tab_bond_4}). The calculated adiabatic energy gaps are 
displayed in Tab.~\ref{Tab_dft_mine_4}). Again LDA and BP86 are found to (massively) over-stabilize the LS state and 
the adiabatic energy gap turns out unrealistically large.

At variance to the previous cases, PBE0 and B3LYP return now an almost identical adiabatic energy gaps. In fact, the 
B3LYP calculated $\Delta E^{\mathrm{adia}}$ is about 30~meV smaller that the PBE0 one and, therefore, the trend 
observed through the series in Eq.~(\ref{func_series}) is not respected. As we will discuss in detail in the following 
sections, this result might be related to the fact that the energetic of [Fe(CO)$_6$]$^{2+}$ depends largely the correlation 
part of the functionals as well as the exchange part. Finally we observe that the Becke-HH functional incorrectly 
predict a HS ground state, meaning that this includes a too large fraction of HF exchange to account accurately for the 
electronic structure of this ion.
\begin{table}[ht!]\centering
\begin{tabular}{lccc}\hline\hline
System & Functional   &  $\Delta E^\mathrm{ZPE}$ (eV) & $\Delta E^\mathrm{ZPE}$ (cm$^{-1}$)\\ \hline\hline
[Fe(H$_2$O)$_6$]$^{2+}$  & BP86   & $-0.08195$ & $-661$\\
$[$Fe(H$_2$O)$_6$]$^{2+}$  & B3LYP  & $-0.09079$ & $-732$\\
$[$Fe(H$_2$O)$_6$]$^{2+}$  & PBE0  & $-0.09308$ & $-750$\\
$[$Fe(H$_2$O)$_6$]$^{2+}$  & HH  & $-0.10272$ & \\
$[$Fe(NH$_3$)$_6$]$^{2+}$ & BP86   & $-0.17413$ & $-1404$\\
$[$Fe(NH$_3$)$_6$]$^{2+}$  & B3LYP  & $-0.16099$ & $-1298$  \\
$[$Fe(NH$_3$)$_6$]$^{2+}$ & PBE0   & $-0.17636$ &  $-1422$\\
$[$Fe(NH$_3$)$_6$]$^{2+}$  &HH  &$-0.16071$ & $-0.16071$  \\
$[$Fe(NCH)$_6$]$^{2+}$  & BP86&  $-0.16593$ & $-1338$\\
$[$Fe(NCH)$_6$]$^{2+}$  & B3LYP & $-0.15367$ & $-1239$\\
$[$Fe(NCH)$_6$]$^{2+}$  & PBE0 & $-0.13846$ & $-1116$\\
$[$Fe(NCH)$_6$]$^{2+}$  & HH & $-0.14487$ & \\
$[$Fe(CO)$_6$]$^{2+}$  & BP86 &  $-0.20800$ & $-1677$\\
$[$Fe(CO)$_6$]$^{2+}$  & B3LYP & $-0.19894$& $-1604$\\
$[$Fe(CO)$_6$]$^{2+}$  & PBE0 & $-0.21157$ & $-1706$\\
$[$Fe(CO)$_6$]$^{2+}$  & HH & $0.21799$ & \\ \hline\hline
\end{tabular}
\caption{Energy difference between the phonon zero point energy of the HS and LS state calculated with the 
various functionals employed in this work (only results for the basis set C are shown).}\label{Tab_phonon_diff}
\end{table}

\subsection*{Zero point phononic energies}
So far we have focused only on the adiabatic energy gaps. However the expression for the internal energy difference, 
Eq.~(\ref{delta_E_int}), contains also a contribution coming from the phonon zero point energies. Table~\ref{Tab_phonon_diff} 
displays $\Delta E^\mathrm{ZPE}$, calculated by using the various functionals (the results are shown only for the basis set 
C). $\Delta E^\mathrm{ZPE}$ is found to be always negative (i.e. the zero point energy of the HS state is lower than that of 
the LS one) reflecting the weaker Fe-ligand bond of the HS configuration. Corrections to the total energy of the two states 
then always tend to stablilise the HS.

In contrast to the adiabatic energy gap,  $\Delta E^\mathrm{ZPE}$ is found to be almost functional independent. 
Indeed, for a given system, the difference between two values of $\Delta E^\mathrm{ZPE}$ obtained with two different 
functionals, are never larger than 15~meV. This demonstrates that the curvature of the PESs is usually very well reproduced 
by every functional. Therefore the spread in the predicted values of $\Delta E^{\mathrm{adia}}$ must arise from the relative 
shift of the PES of one spin state with respect to that of the other. This observation is consistent with the results by Zein 
et al. \cite{Zein}, which indicates that the DOGs, defined by Eq.~(\ref{DOG}), do not depend on the choice of functional. 
\begin{table}[h!]\centering
\scalebox{0.82}{\begin{tabular}{lcccc}\hline\hline
Details geom. opt.   & $\Delta\tau$ (a.u.)&  $E_\mathrm{LS}$(eV) & $E_\mathrm{HS}$(eV) & $\Delta E^{\mathrm{adia}}$ (eV)\\ \hline\hline
BP86 (Basis C)  &  $0.0125$  & $-6127.211(9)$  &  $-6129.720(8)$& $-2.51(1)$\\
BP86 (Basis C)  &  $0.005$  & $-6127.218(9)$  & $-6129.90(2)$& $-2.65(1)$\\
BP86 (Basis C)  &  $0.001$  & $-6127.54(9)$ & $ -6130.19(4)$ & $-2.65(9)$\\
B3LYP (Basis C)  & $0.0125$ &  $-6127.09(2)$ & $ -6129.74(1)$ &$-2.65(2)$\\
B3LYP (Basis C)  &  $0.005$  & $-6127.36(1)$  & $-6129.89(2)$&$-2.54(1)$\\
B3LYP (Basis C)  &  $0.002$  &  $-6127.44(3)$  & $-6130.01(2)$&$-2.57(4)$\\
B3LYP (Basis C)  &  $0.001$  &  $-6127.5(1)$ &  $ -6130.10(2)$&$-2.6(1)$\\
PBE0 (Basis C)  &  $0.125$  & $-6127.220(9)$ &   $-6129.804(8)$ & $2.58(1)$\\
PBE0 (Basis C)  &  $0.005$  &  $-6127.44(2)$ & $-6129.94(2)$&$-2.50(3)$\\
PBE0 (Basis C)  &  $0.001$  & $-6127.66(6)$ &  $-6130.18(4)$& $-2.52(7)$\\ \hline\hline
\end{tabular}}
\caption{DMC total energy for the LS state, the HS state and the adiabatic energy gap of the Fe(H$_2$O)$_6$]$^{2+}$ ion. 
The molecular structures were optimized by DFT using the various functionals and basis sets listed in the first column. 
The time-steps chosen for the DMC simulation are also indicated. Differences in energy are well converged for 
$\Delta\tau=0.005$ a.u.}
\label{Tab_QMC_1}
\end{table}
\begin{table}[h!]\centering
\scalebox{0.82}{\begin{tabular}{lcccc}\hline\hline
Details geom. opt.   & $\Delta\tau$ (a.u.)&  $E_\mathrm{LS}$(eV) & $E_\mathrm{HS}$(eV)& $\Delta E^{\mathrm{adia}}$ (eV)\\ \hline\hline
LDA (Basis C)  & $0.0125$ & $-5234.92(1)$  &  $-5236.93(1)$ & $-2.01(1)$\\
LDA (Basis C)  &  $0.005$  & $-5235.33(2)$   & $-5237.17(1)$ & $-1.84(2)$\\
LDA (Basis C)  &  $0.001$  &  $-5235.69(5)$  &  $-5237.36(5)$ & $-1.67(7)$\\
BP86 (Basis C)  & $0.0125$ &  $-5235.56(1)$ &  $-5237.162(9)$ &$-1.60(1)$\\
BP86 (Basis C)  &  $0.005$  &  $-5235.78(1)$  & $-5237.37(1)$& $-1.58(1)$\\
BP86 (Basis C)  &  $0.001$  &  $-5235.98(3)$ &  $-5237.55(5)$ & $-1.57(5)$\\
B3LYP (Basis C)  & $0.0125$ &  $-5235.516(9)$ &  $-5237.15(1)$ &$-1.63(1)$\\
B3LYP (Basis C)  &  $0.005$  &  $-5235.77(1)$  & $-5237.36(1)$& $-1.59(1)$\\
B3LYP (Basis C)  &  $0.001$  &  $-5236.01(3)$ &  $-5237.59(4)$ &$-1.58(5)$\\
PBE0 (Basis B)  & $0.0125$  &  $-5235.60(1)$  &  $-5237.21(1)$&$-1.61(1)$\\
PBE0 (Basis B)  & $0.005$  &  $-5235.89(2)$  & $-5237.40(2)$ &$-1.51(2)$\\
PBE0 (Basis B)  & $0.001$  &  $-5236.14(3)$  & $-5237.67(9)$&$-1.53(9)$\\
PBE0 (Basis C)  & $0.0125$ &  $ -5235.57(1)$ &  $ -5237.133(8)$ &$-1.56(1)$\\
PBE0 (Basis C)  &  $0.005$  &  $-5235.88(2)$  & $ -5237.37(1)$& $-1.49(1)$\\
PBE0 (Basis C)  &  $0.001$  &  $-5236.10(3)$ &  $ -5237.60(2)$& $-1.50(4)$\\ \hline\hline
\end{tabular}}
\caption{DMC total energy for the LS state, the HS state and the adiabatic energy gap of the [Fe(NH$_3$)$_6$]$^{2+}$ ion. 
The molecular structures were optimized by DFT using the various functionals and basis sets listed in the first column. 
The time-steps chosen for the DMC simulation are also indicated. Differences in energy are well converged for 
$\Delta\tau=0.005$ a.u.}
\label{Tab_QMC_2}
\end{table}
\begin{table}[h!]\centering
\scalebox{0.82}{\begin{tabular}{lcccc}\hline\hline
Details geom. opt.   & $\Delta\tau$ (a.u)&  $E_\mathrm{LS}$(eV) & $E_\mathrm{HS}$(eV) &$\Delta E^{\mathrm{adia}}$ (eV)\\ \hline\hline
BP86 (Basis C)  &  $0.0125$  & $-5957.57(1)$ & $-5959.30(1)$&$-1.73(2)$\\
BP86 (Basis C)  & $0.005$ & $-5957.57(2)$  & $-5959.32(2)$&$-1.75(3)$\\
B3LYP (Basis C)  & $0.0125$ &  $-5957.94(1)$ & $-5959.32(1)$& $-1.38(2)$\\
B3LYP (Basis C)  &  $0.005$ &  $-5957.96(2)$  & $-5959.33(2)$&$-1.37(3)$\\
PBE0 (Basis C)  &  $0.0125$  & $-5957.94(1)$ &  $-5959.291(9)$ &$-1.35(1)$\\
PBE0 (Basis C)  &  $0.005$  & $-5957.95(3)$ & $-5959.30(1)$ &$-1.35(3)$\\ \hline\hline
\end{tabular}}
\caption{DMC total energy for the LS state, the HS state and the adiabatic energy gap of the [Fe(NCH)$_6$]$^{2+}$ ion. 
The molecular structures were optimized by DFT using the various functionals and basis sets listed in the first column. 
The time-steps chosen for the DMC simulation are also indicated. Differences in energy are well converged for 
$\Delta\tau=0.0125$ a.u.}
\label{Tab_QMC_3}
\end{table}
\begin{table}[h!]\centering
\scalebox{0.82}{\begin{tabular}{lcccc}\hline\hline
Details DFT geom. opt.   & $\Delta\tau$ (a.u)&  $E_\mathrm{LS}$(eV) & $E_\mathrm{HS}$(eV) &$\Delta E^{\mathrm{adia}}$ (eV)\\ \hline\hline
B3LYP (Basis C)  & $0.0125$ &  $-6850.97(2)$ & $-6850.64(2)$ & $0.33(3)$\\
B3LYP (Basis C)  & $0.005$ &  $-6850.82(2)$ &  $-6850.45(2)$ & $0.37(3)$\\ \hline\hline
\end{tabular}}
\caption{DMC total energy for the LS state, the HS state and the adiabatic energy gap of the [Fe(CO)$_6$]$^{2+}$ ion. 
The molecular structures were optimized by DFT using the functionals and the basis sets listed in the first column. The 
time-steps chosen for the DMC simulation are also indicated. Differences in energy are well converged for 
$\Delta\tau=0.0125$ a.u.}
\label{Tab_QMC_4}
\end{table}

\section{DMC results and discussion} 
The tables \ref{Tab_QMC_1}, \ref{Tab_QMC_2}, \ref{Tab_QMC_3} and \ref{Tab_QMC_4} display the DMC total 
energies \cite{foot4} of the four ions [Fe(H$_2$O)$_6$]$^{2+}$, [Fe(NH$_3$)$_6$]$^{2+}$, [Fe(NCH)$_6$]$^{2+}$ and 
[Fe(CO)$_6$]$^{2+}$ in both the HS and LS states. The molecular geometries were obtained by DFT optimization (both 
the functional and the basis set used are indicated in the first column on the left-hand side). Unfortunately in most cases 
the DMC energies have a statistical error not small enough to firmly establish which functional returns the lowest energy 
structure of a given complex (only the LDA molecular structures have systematically higher energies, but this is not surprising 
since the analysis of the phonon modes revealed that these structures are not even associated to a stable minimum of the 
LDA total energy). 
\begin{figure}[ht!]\centering
\centering \includegraphics[scale=0.33,clip=true]{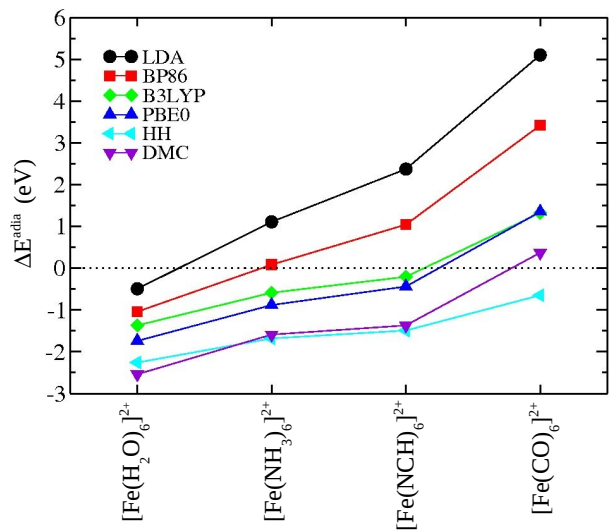}
\caption{Adiabatic energy gaps calculated with DFT and DMC. The DFT results were obtained with the functionals 
indicated in the legend and the basis set C. The DMC results were obtained for the structures optimized with B3LYP 
(basis set C) and with time-steps $\Delta \tau=0.005$ a.u. (the error bars are smaller than the symbols).}\label{fig_delta_E_adia}
\end{figure}

In contrast, the adiabatic energy gaps are calculated with great confidence and they are listed in the right-most column of 
Tabs.~\ref{Tab_QMC_1}, \ref{Tab_QMC_2}, \ref{Tab_QMC_3} and \ref{Tab_QMC_4}. An analysis of these results can be 
carried out by looking at Fig.~\ref{fig_delta_E_adia}, where we present $\Delta E^{\mathrm{adia}}$ calculated with both 
DFT and DMC for all the four ions. The systematic up-shift of the LDA and BP86 values with respect to the DMC ones 
reflects the massive artificial over-stabilization of the LS state (this shift can be as large as few eV). Notably, LDA and 
BP86 incorrectly return a LS ground state for the ions [Fe(NH$_3$)$_6$]$^{2+}$ and [Fe(NCH)$_6$]$^{2+}$.
\begin{figure}[ht!]\centering
\centering \includegraphics[scale=0.33,clip=true]{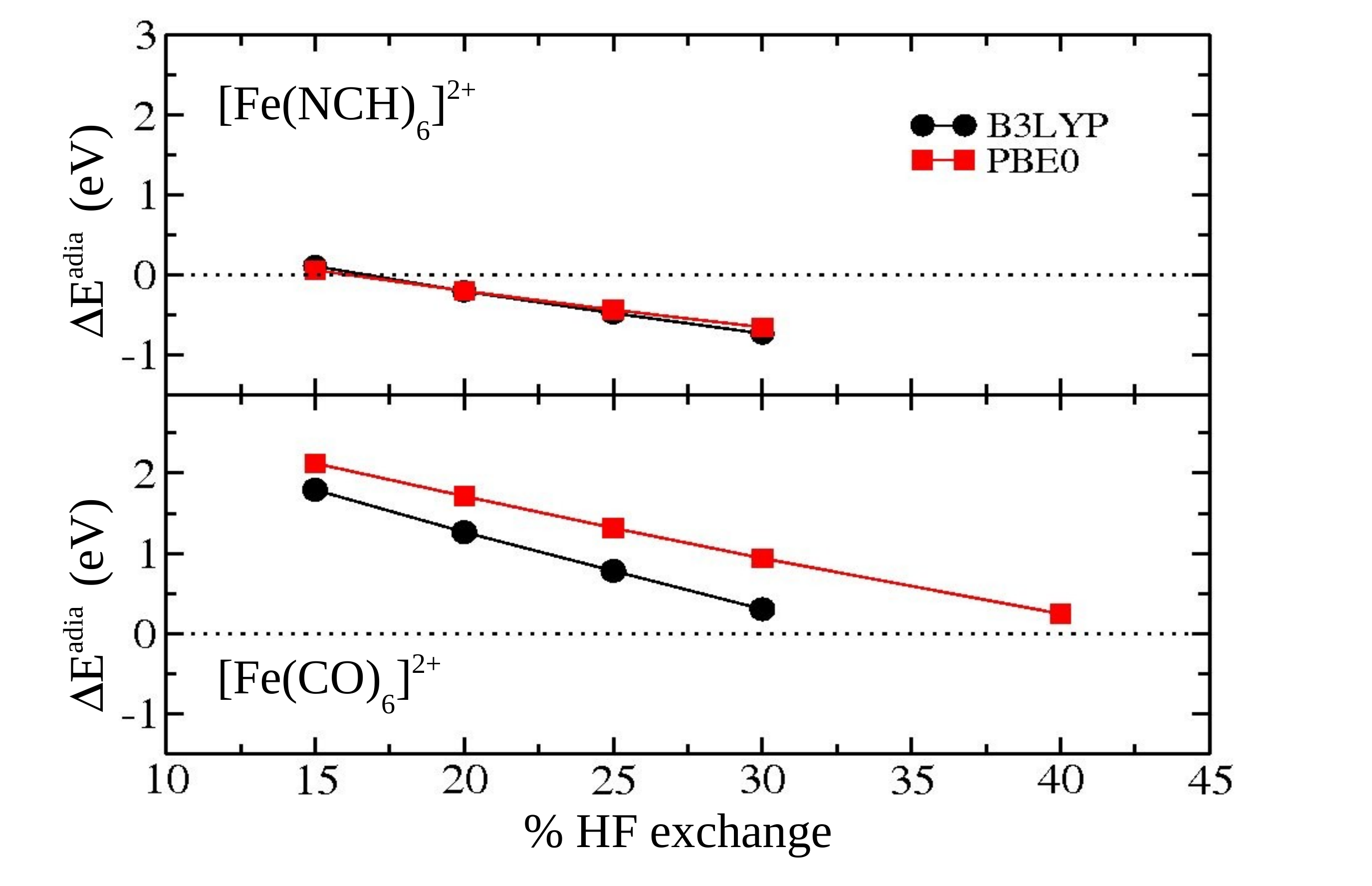}
\caption{Adiabatic energy gaps versus the fraction of HF exchange included in the hybrid functionals B3LYP and PBE0 for 
[Fe(NCH)$_6$]$^{2+}$ (upper pannel) and [Fe(CO)$_6$]$^{2+}$ (lower pannel). The basis set C was used.}\label{fig_delta_E_adia_versus_exx}
\end{figure}

B3LYP and PBE0 provide slightly improved results. Their values for $\Delta E^{\mathrm{adia}}$ lie systematically below 
those computed with BP86 and the ground state spin is correctly predicted for all ions. However, unfortunately, the 
quantitative agreement with DMC is still far from being reached as the PBE0 and the DMC results differ by about 0.6~eV 
(in the best case). Nevertheless hybrid functionals calculate correctly the relative ligand-field strength of the three HS ions 
[Fe(H$_2$O)$_6$]$^{2+}$, [Fe(NH$_3$)$_6$]$^{2+}$ and [Fe(NCH)$_6$]$^{2+}$. In fact, although the B3LYP and PBE0 
results appear shifted vertically according to the fraction of HF exchange included in the functional, the relative 
$\Delta E^{\mathrm{adia}}$ of two complexes is well reproduced. In contrast, the results for [Fe(CO)$_6$]$^{2+}$ do not 
show the same trend. The PBE0 adiabatic energy gap lie slightly above (by about 30~meV) the B3LYP one, despite the 
fraction of HF exchange being larger in PBE0 than in B3LYP. This indicates that the exchange and correlation energies 
have different relative importance for the HS and the LS compounds. In order to better understand this important 
observation, we have calculated the adiabatic energy gap of [Fe(NCH)$_6$]$^{2+}$ and [Fe(CO)$_6$]$^{2+}$ after 
changing the fraction of HF exchange in both B3LYP and PBE0 (see Fig. \ref{fig_delta_E_adia_versus_exx}). 

On the one hand, PBE0 and B3LYP give very similar results for [Fe(NCH)$_6$]$^{2+}$, regardless of the different 
local-exchange and correlation energy. Therefore the calculated adiabatic energy gaps depend mainly on the fraction 
of HF exchange (and this dependence is almost linear). This indicates that the correlation contribution to the total energy 
is well described with (semi-)local functionals and the failures in predicting $\Delta E^{\mathrm{adia}}$ could be entirely 
ascribed to the underestimation of the exchange energy. In addition, by fitting the data, we also conclude that about 
$50\%$ of HF exchange is required to achieve a fair agreement between the DFT and the DMC adiabatic energy gaps. 
Hence, the Becke-HH functional is found to provide quite satisfactory results (see Fig. \ref{fig_delta_E_adia}). 

On the other hand, the B3LYP-calculated $\Delta E^{\mathrm{adia}}$ of [Fe(CO)$_6$]$^{2+}$ is systematically 
down-shifted with respect to the PBE0 value calculated with the same amount of HF exchange. Therefore for this ion the 
results depend drastically on the correlation as well as the exchange part of the density functional. In addition, we note that 
about $30\%$ and $40\%$ of HF exchange, respectively in B3LYP and PBE0, is required to reproduce the DMC gaps and 
that the HH functional incorrectly describes [Fe(CO)$_6$]$^{2+}$ as a HS ion (see Fig. \ref{fig_delta_E_adia}).

The DFT performances for the three HS ions and [Fe(CO)$_6$]$^{2+}$ is related to the different nature of their ground 
state wave-function. In fact, this was found to have a much more pronounced multi-configurational character in [Fe(CO)$_6$]$^{2+}$ 
than in all the other complexes \cite{Domingo}, reflecting the increase in the covalency of the metal-ligand 
bonds \cite{Pierloot,Domingo}. Based on this observation, one can reasonably argue that, for the HS complexes the local-part of the 
exchange-correlation functional is able to capture most of the correlation energy, while the fraction of HF exchange effectively cures 
the LDA underestimation of the exchange. Thus hybrid functionals with ``enough'' HF exchange are found to systematically return 
quite satisfactory results. Furthermore, as the failures of standard GGA functionals seem mostly related to the shortcomings in 
the description of the exchange energy, recent GGA functionals, constructed in order to tackle this issue, can out-perform. For 
example, the OLYP functional, whose exchange part (OPTX) is parametrized to reproduce the Hartree-Fock exchange for 
atoms \cite{olyp}, predicts values for the adiabatic energy gap of [Fe(H$_2$O)$_6$]$^{2+}$ and [Fe(NH$_3$)$_6$]$^{2+}$, 
which compare well with those computed with either B3LYP or PBE0 \cite{Pierloot2}. Unfortuantelly, however, OLYP is not as 
accurate as the hybrids for predicting geometry optimizations and bond-lengths \cite{Pierloot2}. Another of such GGA functionals, which was found 
to perform at the level of the hybrids (if not even better) \cite{Fouqueau_2}, is the HCTH407 \cite{HCTH}. Our own results are 
then listed in Tab. \ref{Tab_gga}. The massive improvement, that OLYP and HCTH407 have achieved, over BP86, is evident in the case of 
Fe(H$_2$O)$_6$]$^{2+}$, [Fe(NH$_3$)$_6$]$^{2+}$ and [Fe(NCH)$_6$]$^{2+}$. 
\begin{table}[ht!]\centering
\scalebox{0.9}{\begin{tabular}{lcccc}\hline\hline
System & Functional   & Basis set & $\Delta E^{\mathrm{adia}}$ (cm$^{-1}$)& $\Delta E^{\mathrm{adia}}$ (eV) \\ \hline\hline
[Fe(H$_2$O)$_6$]$^{2+}$&OLYP & C & $-15953$& $-1.9780$\\
$[$Fe(H$_2$O)$_6$]$^{2+}$&HCTH407 & C & $-19315$ & $-2.3947$\\
$[$Fe(NH$_3$)$_6$]$^{2+}$&OLYP & C & $-7338$ & $-0.9099$\\
$[$Fe(NH$_3$)$_6$]$^{2+}$&HCTH407 & C & $-9942$ & $-1.2327$\\
$[$Fe(NCH)$_6$]$^{2+}$&OLYP & C & $525$ & $0.06510$\\
$[$Fe(NCH)$_6$]$^{2+}$&HCTH407 & C & $-3650$ & $-0.4526$\\
$[$Fe(CO)$_6$]$^{2+}$&OLYP & C & $21313$ & $2.6425$\\
$[$Fe(CO)$_6$]$^{2+}$&HCTH407 & C & $17097$ & $2.1198$\\ \hline\hline
\end{tabular}}
\caption{Adiabatic energy gap, $\Delta E^{\mathrm{adia}}$, for the four ions calculated with the OLYP and HCTH407 functionals (the basis sets C was used).}\label{Tab_gga}
\end{table}

In contrast, one can question whether [Fe(CO)$_6$]$^{2+}$ can be described at all by the single-determinant picture 
provided by DFT. In principle, the multiconfigurational nature of a wave-function can be described by GGA functionals. 
In fact the GGA exchange roughly mimics the non-dynamical correlation (in addition to the proper exchange) \cite{Gritsenko}. 
In practice, however, no DFT flavour investigated here has proven fairly accurate for the energetic of the ion 
[Fe(CO)$_6$]$^{2+}$.  

\begin{table}[ht!]\centering
\scalebox{0.82}{\begin{tabular}{lccc}\hline\hline
Method   & Reference  &  $\Delta E^{\mathrm{adia}}$ (cm$^{-1}$)& $\Delta E^{\mathrm{adia}}$ (eV) \\ \hline\hline
CASSCF(6,5) & \cite{Fouqueau_1}& $-23125$ & $-2.86714$\\
CASSCF(12,10)& \cite{Fouqueau_1}& $-21180$ & $-2.62599$\\
corr-CASSCF(12,10)& \cite{Fouqueau_1}& $-17892$ & $-2.21833$\\
CASPT2(6,5) & \cite{Fouqueau_1}& $-21610$ & $-2.6793$\\
CASPT2(12,10)& \cite{Fouqueau_1}& $-16185$ & $-2.00668$\\
corr-CASPT2(12,10)& \cite{Fouqueau_1}& $-12347$ & $-1.53083$\\
SORCI& \cite{Fouqueau_1}& $-13360$ & $-1.65643$\\
CASPT2(10,12) &\cite{Pierloot}&  $-16307$ & $-2.02181$\\ \hline\hline
\end{tabular}}
\caption{The adiabatic energy gap for [Fe(H$_2$O)$_6$]$^{2+}$ calculated by using various wave-function methods 
(reference to the literature is given in the second column). The values labelled as corr-CASSCF and corr-CASPT 
denote respectively the CASSCF and CASPT values after having applied an empirical correction of the order of 
$4000\mathrm{cm}^{-1}$ (see main text). Pierloot et al. \cite{Pierloot} provides an additional long list of results obtained 
by using different basis sets, geometries and symmetries. Here we report only the value that these authors indicate 
as the ``best''.}\label{Tab_cation_1_WF}
\end{table}
\begin{table}[ht!]\centering
\scalebox{0.82}{\begin{tabular}{lccc}\hline\hline
Method   & Reference  &  $\Delta E^{\mathrm{adia}}$ (cm$^{-1}$)& $\Delta E^{\mathrm{adia}}$ (eV) \\ \hline\hline
CASSCF(12,10)& \cite{Fouqueau_2}& $-20630$ & $-2.55779$\\
corr-CASSCF(12,10)& \cite{Fouqueau_2}& $-16792$ & $-2.08194$\\
CASPT2(12,10)& \cite{Fouqueau_2}& $-12963$ & $-1.60721$\\
corr-CASPT2(12,10)& \cite{Fouqueau_2}& $-9125$ & $-1.13136$\\
SORCI& \cite{Fouqueau_2}& $-10390/-11250$ & $-1.2882/-1.39482$\\
CASPT2(12,10)& \cite{Pierloot}& $-7094$ & $-0.879544$\\ \hline\hline
\end{tabular}}
\caption{Adiabatic energy gaps for [Fe(NH$_3$)$_6$]$^{2+}$ calculated by using various wave-function methods 
(reference to the literature is given in the second column). The values labelled as corr-CASSCF and corr-CASPT 
denote respectively the CASSCF and CASPT values after having applied the empirical correction of the order 
of $4000\mathrm{cm}^{-1}$ (see main text). }\label{Tab_cation_2_WF}
\label{Tab_cation_2}
\end{table}

Finally we compare our DMC results to those obtained with wave-function based methods. Fouqueau 
et al. \cite{Fouqueau_1,Fouqueau_2} carried out several calculations for the adiabatic energy gap of the ions 
[Fe(H$_2$O)$_6$]$^{2+}$ and [Fe(NH$_3$)$_6$]$^{2+}$ by using the complete active space self-consistent field 
(CASSCF) method with second order perturbative corrections (CASPT2). Some of the results are summarized in 
Tabs.~\ref{Tab_cation_1_WF} and \ref{Tab_cation_2_WF}. As observed by the authors themselves and in Ref.~\cite{Pierloot}, 
these calculations suffer the drawback of having been carried out with an insufficient Fe basis set. As such they are
affected by a systematic error, which can be estimated by considering the $^5D-^1I$ splitting of the free Fe$^{2+}$ ion. 
An empirical correction of the order of $4000\,\mathrm{cm}^{-1}$ was then introduced. In the same works, results 
obtained by spectroscopy-oriented configuration-interaction (SORCI), were also reported. These were stated not to 
require any empirical correction.
A second set of results is provided by Pierloot et al. \cite{Pierloot}, who performed calculations with basis sets of larger size. 
For [Fe(NH$_3$)$_6$]$^{2+}$, their best CASPT2 adiabatic gap agrees fairly well with the corrected-CASPT2 and SORCI 
results. However for [Fe(H$_2$O)$_6$]$^{2+}$, they found a significantly larger (in absolute value) $\Delta E^{\mathrm{adia}}$.

By analyzing the data in Tabs.~\ref{Tab_cation_1_WF} and \ref{Tab_cation_2_WF}, we note that the adiabatic energy 
gaps calculated with CASPT2 by Fouqueau et al. \cite{Fouqueau_1,Fouqueau_2} agree fairly well with our DMC ones 
(in particular for [Fe(NH$_3$)$_6$]$^{2+}$). In contrast, the empirical corrections worsen the agreement and the SORCI 
results do not agree quantitatively with ours. Although we have not achieved yet a complete understanding of these 
differences, we argue that they may originate from the large dependence of the CASSCF/CAST2 results on the 
basis sets and on the orbitals included in the active space. Furthermore wave-function based methods do not describe 
dynamic electronic correlations (although partial corrections are provided by the second order perturbation theory). The DMC 
energies, in turns, might depend on the choice of the trial wave-function introduced to impose the fixed-node approximation 
\cite{Foulkes}. A thorough analysis on the sources of potential errors in DMC is currently on the way.

\section{Conclusions}
In this work, we have assessed the performances of several popular exchange-correlation functionals in describing 
various Fe$^{2+}$ complexes. As DFT results can not be easily related to experiments (at least without accounting for 
environmental and finite-temperature effects), we have performed accurate DMC calculations, which provide a solid
theoretical benchmark for the theory. The DFT and DMC results, both obtained within the theoretical framework of 
the adiabatic approximation, could be then directly compared.

The LDA and the standard GGA functionals drastically over-stabilize the LS state. Although the accuracy of the DFT calculations 
increases when hybrid functionals are employed, the most popular ones, B3LYP and PBE0, provide results, which are still 
quantitatively unsatisfactory. In the case of HS ions, a fair agreement between the DFT and the DMC adiabatic gaps is achieved 
only by using about $50\%$ of HF exchange. In contrast, a lower fraction of HF exchange (between $30\%$ and $40\%$) is 
required for [Fe(CO)$_6$]$^{2+}$. This difference might be related to the diverse nature of the ground state wave-function for 
the HS and LS ions. Therefore, unfortunately, we have to conclude that there is not yet a ``universal'' functional able to correctly 
describe the energetics of every Fe$^{2+}$ complex.

Finally, by analyzing zero-point phonon energies, we have demonstrated that the shape of the PESs is well described by 
every functional considered. Therefore, as already pointed out by Zein et al. \cite{Zein}, the failures of DFT in calculating 
the adiabatic energy gaps must be ascribed to a shift of the PES of the LS state with respect to that of the HS state.

\section{Acknowledgments}
The authors thank N. Baadji for useful discussions. This work is sponsored by the EU-FP7 program (HINTS project). 
Computational resources have been provided by the Trinity Center for High Performance Computing (TCHPC) and 
by the Irish Center for High End Computing (ICHEC).


\begin{thebibliography}{100}
\bibitem{foot1}We use here the standard SC terminology often employed in literature. In this work, by gap 
we will always mean either the adiabatic energy difference, Eq.~(\ref{delta_E_adia}), or the vertical energy 
difference, Eq.~(\ref{vert_gap}). These must not be confused with the actual electronic gap of a molecule 
defined as the difference between the ionization potential and the electron affinity.
\bibitem{foot2}In contrast, we have found that some more recent GGA functionals, such as the OLYP, return metal-ligand bonds longer than those predicted by hybrids.
\bibitem{foot4}Note that the notation $E=-5957.57(1)$ eV stands for $E=(-5957.57\pm 0.01)$~eV.
\end{thebibliography}
\end{document}